\documentclass[preprint2,twoside]{hwo}

\usepackage{graphicx}
\usepackage{lipsum}
\usepackage{tabularx}
\usepackage{array}
\usepackage[margin=1in]{geometry}

\newcommand{\fesc}{$f_{esc}^{LyC}$}
\newcommand{\ro}{$\rho_{UV}$}
\newcommand{\xion}{$\xi_{\rm ION}$}
\newcommand{\nion}{$n_{\rm ION}$}
\newcommand{\Muv}{$M_{UV}$}
\newcommand{\fLyC}{$f_{LyC}$}

\input{hwo.h}

\begin{document}

\title{\textbf{\LARGE Identifying the contributors to cosmic reionization with the Habitable World Observatory:
LyC escape calibration in faint galaxies}}


\author {\textbf{\large Annalisa Citro $^{1}$, Cody. A. Carr $^{2}$, Yumi Choi $^{3}$, Sophia. R. Flury $^{4,11}$, Matthew. J. Hayes $^{5}$, Anne Jaskot $^{6}$, Gagandeep Kaur $^{7}$, Alexandra Le Reste $^{1}$, Matilde Mingozzi $^{8}$, Themiya Nanayakkara$^{9}$, Sally Oey $^{10}$, Claudia. M. Scarlata $^{1}$}}

\affil{$^1$\small\it Minnesota Institute for Astrophysics, School of Physics and Astronomy, University of Minnesota, 316 Church Street SE, Minneapolis, MN 55455, USA}

\affil{$^2$\small\it Center for Cosmology and Computational Astrophysics, Institute for Advanced Study in Physics, Zhejiang University, Hangzhou 310058, People’s Republic of China}

\affil{$^3$\small\it NSF National Optical-Infrared Astronomy Research Laboratory, 950 North Cherry Avenue, Tucson, AZ 85719, USA}

\affil{$^4$\small\it Department of Astronomy, University of Massachusetts Amherst, Amherst, MA 01002, USA}

\affil{$^5$\small\it Stockholm University, Department of Astronomy and Oskar Klein Centre for Cosmoparticle Physics, AlbaNova University Centre, SE-10691, Stockholm, Sweden}

\affil{$^6$\small\it Department of Astronomy, Williams College, Williamstown, MA 01267, USA}

\affil{$^7$\small\it Department of Astronomy and Oskar Klein Centre for Cosmoparticle Physics, Stockholm University, SE-10691, Stockholm, Sweden}

\affil{$^8$\small\it AURA for ESA, Space Telescope Science Institute, 3700 San Martin Drive, Baltimore, MD 21218, USA}

\affil{$^9$\small\it Centre for Astrophysics and Supercomputing, Swinburne University of Technology, PO Box 218, Hawthorn, VIC 3122, Australia
}

\affil{$^{10}$\small\it Astronomy Department, University of Michigan, 311 West Hall, 1085 S. University Ave., Ann Arbor, MI 48109-1107, USA}

\affil{$^{11}$\small\it Institute for Astronomy, University of Edinburgh, Royal Observatory, Edinburgh, EH9 3HJ, UK}

\author{\footnotesize{\bf Endorsed by:}
Anshuman Acharya (Max Planck Institute for Astrophysics, Germany), Karla Z. Arellano-Cordova (Univserity of Edinburgh, UK), Michelle Berg (Texas Christian University, USA), Filippo D'Ammando (INAF-IRA Bologna, Italy), Sophia R. Flury (Univserity of Edinburgh, UK), Farhan Hasan (STScI, USA), Mason Huberty (University of Minnesota - Twin Cities, USA), Brad Koplitz (Arizona State University, USA), Joshua Oppor (The University of Texas at Austin, USA), Marc Rafelski (STScI, USA), Faraz Nasir Saleem (Egypt Space Agency (EgSA))
}



\begin{abstract}

Investigating how small-scale physical processes shape large-scale astrophysical phenomena is one of the key science themes of the Astro2020 decadal Survey. An example of this interplay is Cosmic Reionization, where ionizing photons from galaxies escaped into the intergalactic medium (IGM), driving its transition from neutral to ionized at $z>6$. Star-forming galaxies are thought to be the dominant sources of reionization. {However, an open question remains as to whether bright or faint star-forming galaxies are the primary reionization contributors.} This depends on the escape fraction \fesc\ -- the fraction of ionizing photons that successfully escape the galaxy's interstellar and circumgalactic medium to reach the IGM. Performing direct measurements of \fesc during the reionization epoch is not feasible, due to the near-zero transmission of the IGM at $z\gtrsim4$. 
However, by calibrating \fesc\ against \textit{indirect indicators} (i.e. galaxy properties that correlate with \fesc) at lower redshifts, we can estimate its value in galaxies during reionization. Currently, \fesc\ calibrations are mostly limited to $\sim 90$ sources brighter $M_{UV} \sim -18$ at $z\sim0.3$. Yet, recent models suggest that \textit{fainter galaxies} ($-19<M_{UV}<-13$) may be the major contributors to reionization. {Our science goal is to extend \fesc\ calibrations via indirect indicators to such faint magnitudes.} With the Habitable Worlds Observatory, we aim to achieve a $\sim$ 20-fold increase in statistical power. Additionally, we plan to reach magnitudes as faint as 1/100 $L^*$ at $z\sim0.1$. Suitable targets will be selected from the upcoming Ultraviolet Explorer (UVEX) survey. Through this science case, we will be able to place robust constraints on the role of faint galaxies in the reionization of the Universe and identify its primary contributors.
  \\
  \\
\end{abstract}

\vspace{2cm}

\section{Science goal}
\label{goal}

\noindent The Astro2020 Decadal Survey has identified “Cosmic Ecosystems” as a key science theme for the current decade. This science theme aims at exploring how small-scale physical processes at work in galactic ecosystems can have an impact on the large-scale behavior of astrophysical systems. Three central questions have been formulated to guide this investigation: 
\begin{itemize}

\item[-]How did the intergalactic medium and the first sources of radiation evolve from cosmic dawn through the epoch of reionization? 

\item[-]How do gas, metals, and dust flow into, through, and out of galaxies?

\item[-]How do supermassive black holes form and how is their growth coupled to the evolution of their host galaxies? 

\end{itemize}

\noindent The current Habitable World Observatory (HWO) science case aims to address point 1).  Specifically, our science question is: \textbf{Which are the main contributors to the reionization of the Universe?}\\

\noindent A key example of the interplay between physical processes on a small and large scale is ionizing radiation. Ionizing photons, commonly referred to as Lyman continuum (LyC) photons, have energies exceeding 13.6 eV (i.e. wavelengths $\lambda$ shorter than 912 \AA) and are generated on small scales by young, massive, O- and early B-type stars, within galaxies. Although much of this radiation is absorbed locally, a fraction can escape the dense galactic environment through outflows or low-density channels in the interstellar medium and leak into the intergalactic medium (IGM). This escape of ionizing photons is what enabled the first generations of stars to drive a large-scale cosmic phase transition at $z\gtrsim6$, known as the Epoch of Reionization (EoR), during which the majority of hydrogen in the universe transitioned from a neutral to an ionized state.  

Exploring this aspect of the small-to-large-scale connection requires addressing a fundamental question: which were the main contributors to this process? 

\vspace{1mm}
Reionization sources are parameterized through the emissivity of ionizing photons per unit time and volume:  

\begin{equation}
n_{\rm ION} = \rho_{UV} * \xi_{\rm ION} * f_{esc}^{LyC}~~,
\end{equation}

Where \ro\  is the volume density of ionizing radiation-producing objects, \xion\ is the intrinsic production rate of ionizing photons per unit non-ionizing UV luminosity,  and \fesc\ is the escape fraction of the produced ionizing photons, i.e. the fraction of ionizing photons that successfully escape the galaxy’s interstellar and circumgalactic medium to reionize the diffuse IGM. By determining \nion\ over cosmic time and across various types of sources, we can identify the major contributors to the reionization process. 

Unfortunately, deriving \ro,  \xion, and \fesc\ present significant challenges. While \ro\ and  \xion\ can be directly derived at the EoR ($z\sim6$), \fesc\ can be measured only up to $z\sim4$, beyond which the transmission of the IGM drops to zero. 
In the pre-JWST era, several studies favored star-forming galaxies as the dominant sources of reionization over active galactic nuclei \citep[e.g,][]{Ouchi+2009, Naidu+2020, Donnan+2023}. However, within the category of star-forming galaxies, a debate has emerged regarding whether faint or bright sources primarily contributed to reionization. Some studies have suggested that the bulk of ionizing photons came from numerous faint galaxies \citep[e.g.,][]{Oesch+2009, Wise+2014, Livermore+2017, Finkelstein+2019}. Conversely, other research suggests that a few bright galaxies were sufficient to reionize the universe \citep[e.g.,][]{Naidu+2020, Mason+2020}. 

Understanding which sources dominated the reionization process is crucial, as it directly influences the timescale over which reionization unfolded. If faint, low-mass galaxies were the primary contributors, reionization likely progressed more gradually over extended periods. In contrast, if bright, massive galaxies played the leading role, the process would have occurred more rapidly. This distinction carries significant implications for galaxy formation and evolution across different mass regimes, as the timing and intensity of reionization can suppress star formation in low-mass halos \citep{Efstathiou1992, Nebrin+2023, Wu&Kravtsov2024}.

JWST is now providing the opportunity to explore the relative contribution of faint vs. bright star forming galaxies directly in the EoR, where \ro\ and \xion\  are observable. Thanks to the magnification provided by gravitational lensing,  EoR studies of \ro\ and \xion\ have been pushed to extremely faint magnitudes ($M_{UV}\sim-13$).  The emerging picture is that both \ro\ and \xion\ increase for decreasing galaxy brightness at  $2 < z < 9$, supporting a scenario where fainter galaxies contribute more to reionization \citep{Atek+2015, Livermore+2017, Emami+2020, Atek+2022, Simmonds+2024}.

Despite the insights provided by JWST, there is still a key missing piece: the escape fraction \fesc, which is out of our reach at $z > 4$. Fortunately, \fesc\ can be calibrated at lower redshifts using galaxy properties that correlate with the likelihood of LyC photons escaping into the IGM. These properties, (such as dust content, gas covering fraction, gas ionization state, see Section \ref{sec:objective})  are referred to as \fesc\ indirect indicators. Once defined at low redshift, these calibrations can then be applied to EoR galaxies to infer their \fesc. 

\begin{figure*}[htbp]
\centering
    \includegraphics[width = \textwidth]{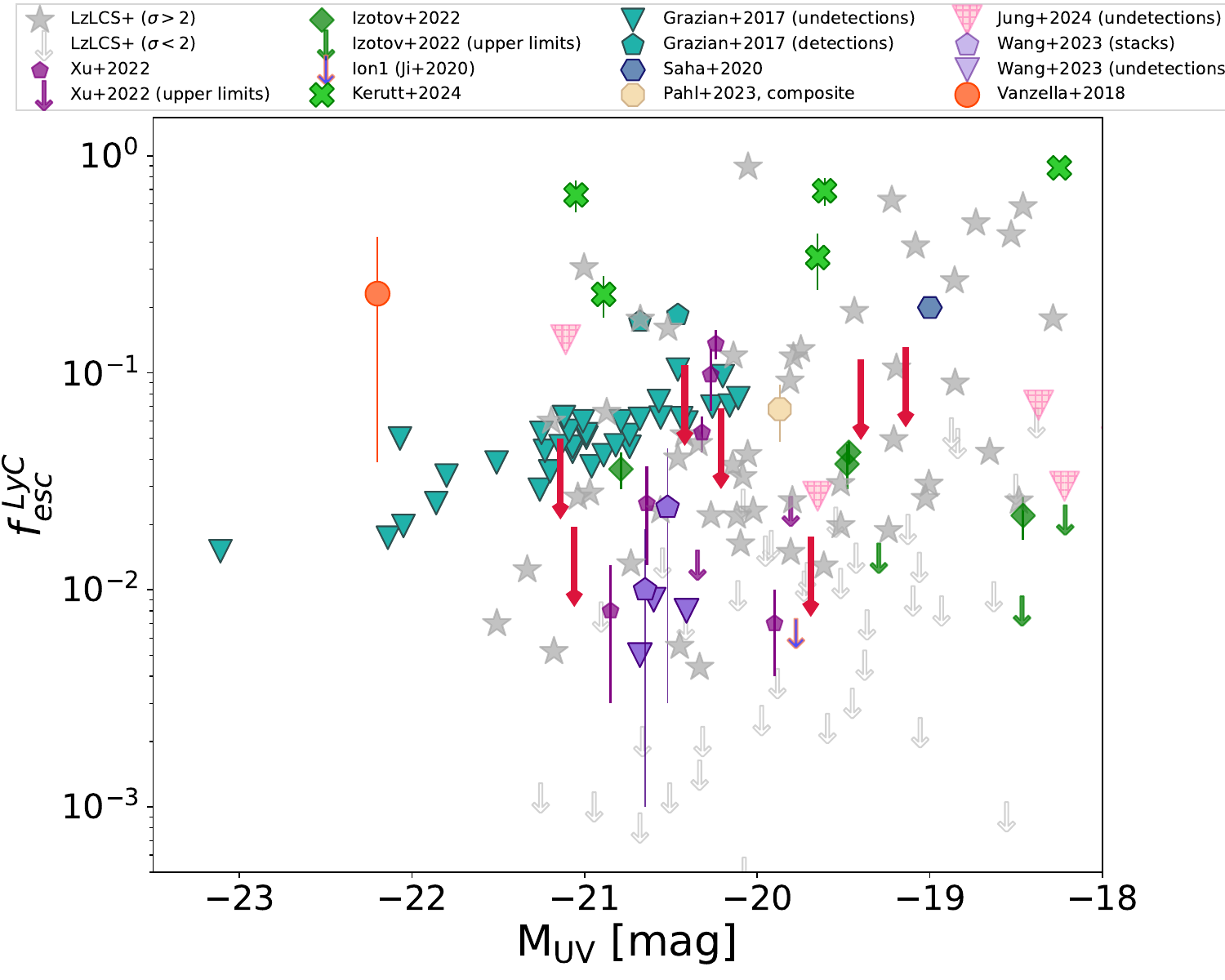}
\caption{Absolute escape fractions as a function of UV magnitude at 1500 $\AA$. local galaxies from the LzLCS sample \citep{Flury+2022a, Flury+2022b}. Red downward pointing arrows are $1\sigma$ upper limits on the absolute escape fraction for the seven BELLS GALLERY galaxies studied in \citet{Citro+2025}. Downward pointing triangles and downward pointing arrows of different colors are non-detections from \citet{Grazian+2016} at $1\sigma$, \citet{Wang+2023} at $2\sigma$, \citet{Jung+2024} at $1\sigma$, and \citet{Ji+2020}. Other symbols are detections from \citet{Grazian+2016, Flury+2022a, Vanzella+2018, Saha+2020, Pahl+2023, Izotov+2022, Xu+2022}, and \citet{Kerutt+2024}, respectively (when relative escape fractions are provided, we transform them into absolute escape fractions, assuming a \citet{Calzetti+2000} attenuation curve and $E(B-V) = 0.1$).}
\label{fig:fesc}
\end{figure*}

At present, \fesc\ indirect indicators are mostly limited to galaxies with \Muv\ brighter than $\sim -18$ (LzLCS+ survey, \citealp{Flury+2022a, Flury+2022b, Izotov+2016a, Izotov+2016b, Izotov+2018a, Izotov+2018b, Izotov+2021, Wang+2023, Grazian+2016, Vanzella+2018, Saha+2020, Mascia+2023, Mingozzi+2024, Jung+2024, Citro+2025}, see Figure \ref{fig:fesc}), with only a few exceptions \citep{Choi+2020}. However, calibrating them at fainter magnitudes is crucial, especially in light of recent works based on models showing that galaxies with $-19 < M_{UV} < -13$ have a higher probability of having a higher escape fraction at $z>6$ \citep{Lin+2024, Mascia+2024}.

\textbf{This science case aims to calibrate indirect indicators of \fesc at the faintest magnitudes, so far barely explored, \Muv $>-18$. By combining these \fesc\ calibrations with measurements of \ro\ and \xion\ in faint galaxies at $z > 6$, we will place robust constraints on the contributors, timeline, and history of cosmic reionization.}

\section{Science objective}
\label{sec:objective}

\begin{figure*}[h]
\centering
    \includegraphics[width = \textwidth, trim={0em, 7em, 0em, 0em}]{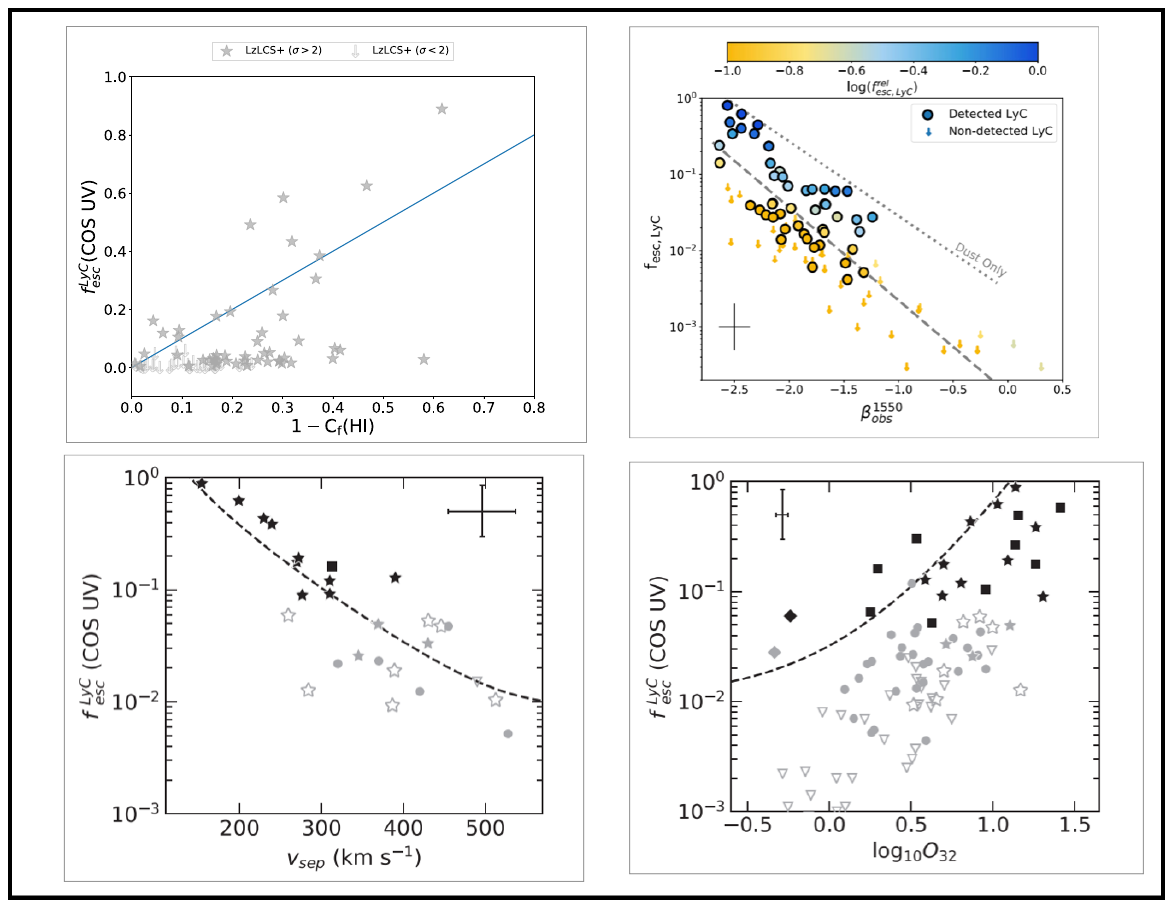}
\caption{Collection of state-of-the art studies about the connection between \fesc\ and indirect indicators. Upper left panel: \fesc\ increases for decreasing covering fraction $C_{f}(HI)$ – or for increasing $1-C_{f}(HI)$. Upper right panel: \fesc\ increased for steeper UV beta slope (lower dust attenuation). Lower left panel: \fesc\ increases for lower separations of the Ly$\alpha$ peaks ($v_{sep}$). Lower right panel: \fesc\ increases for increasing [\ion{O}{3}]/[\ion{O}{2}] ratio.}
\label{fig:fesc2}
\end{figure*}

\noindent \textbf{Derive the escape fraction \fesc\ of ionizing photons in low mass, faint galaxies (down to \Muv $\sim$ -13) at $z \lesssim 0.1$ and calibrate indirect indicators on the derived \fesc.}

\subsection{LyC escape fraction}
The first crucial step in carrying out this science case is to derive the LyC escape fraction \fesc. Our plan is to directly observe the LyC emission in a galaxy (\fLyC) and the emission at 1500 \AA ($f_{1500}$, which corresponds to wavelengths of non-ionizing radiation) and compute the escape fraction using the following equations:

\begin{equation}
f_{\mathrm{esc,rel}} = \frac{(f_{\mathrm{LyC}}/f_{1500})^{\mathrm{out}}}{(L_{\mathrm{LyC}}/L_{1500})^{\mathrm{int}}}~~,
\end{equation}

where $f_{\mathrm LyC}^{\mathrm out}$ is the LyC flux density per unit frequency in the vicinity of the galaxy, right after escaping the ISM, while $f_{\mathrm 1500}^{\mathrm out}$ is the flux density per unit frequency measured at 1500 \AA\ after passing through the galaxy's ISM. These two quantities are related to the observed fluxes (i.e. the fluxes that survive the journey through the IGM to reach us) as follows: 

\begin{equation}
\Big(\frac{f_{\mathrm LyC}}{f_{1500}}\Big)^{\mathrm out} = \Big( \frac{f_{\mathrm LyC}}{f_{1500}} \Big)^{\mathrm obs} \times e^{\tau_{\rm IGM}(\mathrm LyC)}~~,
\end{equation}

where $e^{\tau_{\rm IGM}(\mathrm LyC)}$ is the optical depth of LyC photons through the IGM along the line of sight to that galaxy \footnote{Given that the IGM is mostly ionized at low redshifts, which is what this science project aims to observe, we do not expect the IGM correction to affect our results significantly.}. We will constrain the intrinsic $(f_{\mathrm LyC}/f_{1500})^{\mathrm {int}}$ using fits to the continuum with individual Binary Population and Spectral synthesis (BPASS, \citealp{Eldridge+2017}) or STARBURST99 \citep{Leitherer+1999} templates of young galaxies (which are the ones expected to contribute more to \fesc, see below). We will determine the dust attenuation $A_{1500}$ starting from the UV slope of the stellar continuum ($\beta_{1500}^{\mathrm obs}$). We will convert $\beta_{1500}^{\mathrm obs}$ to E(B-V) using the \citet{Reddy+2016} relations. 

\vspace{1mm}
\noindent \underline{Requirements}: To ensure a reliable determination of \fesc, we require that \fLyC and $f_{1500}$ be measured with a significance of at least 3$\sigma$. Our aim is to reach \fesc\ values of 1\% (i.e., apparent LyC ABmag $\sim$ 32.5 for our major progress goal). 
\\

\subsection{Indirect indicators of the LyC escape fraction}
\noindent Given that \fesc\ cannot be observed at $z > 4$ due to the $\sim$ 0 transmission of the IGM, over the last decades researchers have identified specific galaxy properties - referred to as indirect \fesc\ diagnostics - which correlate with the likelihood of LyC escaping the galaxy’s ISM and CGM and can be studied in the local Universe.  These properties are usually encoded in galaxies’ spectral features at rest-UV and rest-optical wavelengths. 

For example, LyC photons can be absorbed by dust or neutral hydrogen within the ISM and CGM of the galaxy. As a result, \fesc\ is likely higher in galaxies with a patchy or low-neutral hydrogen column density and minimal dust. Therefore, absorption lines that trace the neutral hydrogen density, its distribution, and dust content are among the key indirect indicators of LyC escape.
Additionally, stellar feedback is essential for clearing gas and facilitating the escape of LyC photons. High ionization levels and elevated star formation rates (SFR) per unit area (SFR surface density, or $\rm \Sigma SFR$) indicate robust stellar feedback, which can create pathways for LyC photons to escape. Consequently, LyC escape diagnostics include strong feedback tracers, such as high-ionization UV and optical emission lines and high $\rm \Sigma SFR$.

In the following, we provide a more detailed description of which indirect indicators (and associated spectral features) of \fesc\ we propose to focus on.\\

\noindent 1) \textbf{Ionization state}. The ionization state of the ISM in a galaxy can be traced through emission line ratios between lines with different ionization potentials, such as C IV $\lambda$1500/C III] $\lambda$1909 or [O III] $\lambda$5007/[O II] $\lambda$$\lambda$3727,9, or through individual high ionization emission lines, such as He II $\lambda$1640. Specifically, C IV $\lambda$1550 and C III] $\lambda$$\lambda$1907,9 lines are produced by ions in different ionization states, namely C$^{3+}$  and C$^{2+}$.
Since C$^{3+}$  has a higher ionization potential than C$^{2+}$ , a high C IV $\lambda$1550/C III] $\lambda$$\lambda$1906,9  ratio suggests the presence of many C$^{3+}$ ions, and thus a larger number of photons with energies $\gtrsim$ 48 eV. Similarly, the presence of the He II line, produced by He$^{2+}$ ions , indicates the presence of photons with energy higher than $\sim$ 54 eV, and therefore a highly ionized ISM. Also a high [O III] $\lambda$5007/ [O II] $\lambda$3727 ratio implies a high ionization state for the gas.  (see Figure \ref{fig:fesc2}). This condition favors the escape of ionizing photons, which are less likely to be absorbed by neutral hydrogen. 
\\

\noindent 2) \textbf{ISM hydrogen distribution}. Resonant scattering occurs when high-energy photons interact with neutral atoms. During each interaction, the resonant photons are scattered and can gain or lose energy (depending on the interaction's geometry), gradually shifting away from the resonance wavelength. This happens until the photon's energy is sufficiently altered to travel without further scattering.

Resonant lines can serve as indicators of \fesc\ because they require an optically thin gas to escape the galaxy and be observed. In fact, in an optically thick gas, the number of scatterings increases to the point where the resonant photons become trapped, preventing their escape. Over the past few decades, two resonant lines: Ly$\alpha$ (at 1216 \AA) and Mg II (at 2808 \AA), have been recognized as useful indirect indicators of the LyC escape fraction. Because Mg II has a longer wavelength, it is less affected by absorption from the IGM at high redshifts. For this reason, our science case will focus on Mg II. 

Due to scattering and the resulting shift in velocity space, the profile of resonant lines can exhibit two peaks with a dip at the (most absorbed) resonant wavelength. The relative intensity and shift of these peaks provide information on the LyC escape fraction. Specifically, a larger separation of the peaks suggests that the resonant photons experienced more scattering before escaping and that they have traveled through a more optically thick gas. 

\vspace{1mm}
\noindent \underline{Requirements}: The current state of the art for emission line observations in low-mass galaxies is represented by the CLASSY Survey \citep{Berg+2022}. For these sources, \citet{Mingozzi+2022} found continuum levels of $\rm \sim 2-10 \times 10^{-15} erg~s^{-1}~cm^{-2}~A^{-1}$ at $\sim1600$ \AA (which corresponds to the wavelength of the weakest of the emission lines needed to carry out the science objectives 1) and 2), that is, He II $\lambda$ 1640), with S / N = 10. Assuming a minimum equivalent width for the He II line of 3 \AA (compatible with what is found in CLASSY galaxies), this implies a line flux sensitivity of He II of $\rm \sim 6-30 \times 10^{-15}  erg~s^{-1}~cm^{-2}$. Our aim is to detect the continuum at $\sim$ 1600 with S/N ratio $> 10$ Specifically, a major progress would be to increase the S/N in the continuum by 10 times compared to the CLASSY galaxies and therefore detect 10 times fainter He II lines with the same EW. For point 2), in order to detect the velocity structure of the Mg II line, we require a high spectral resolution ($R = 10,000$), which corresponds to $\sim$ 30 $\rm km~s^{-1}$ at the wavelengths of this line. 
\\

\noindent 3) \textbf{Dust attenuation}. FUV bright stellar populations are typically characterized by very blue colors because their spectral energy distributions peak in the extreme-ultraviolet \citep{Leitherer+1999, Raiter+2010, Eldridge+2017}. The blackbody nature of massive stars leads to a very steep negative power-law index at 1500 \AA\ with $\beta_{1500}^{\mathrm int}$ between $-2.8$ and $-2.5$. The non-ionizing FUV light from massive stars propagates within the galaxy, where it is absorbed and attenuated. This dust attenuation flattens the observed UV spectrum and increases the observed values of $\beta_{1500}^{\mathrm obs}$  (see Figure \ref{fig:fesc2}). By understanding the effects of the nebular continuum and its relationship (or lack thereof) with dust attenuation, we can gain insights into the relative contributions of gas and dust in extinguishing the LyC. 
\\

\noindent 4) \textbf{Covering fraction}. The radiative transfer of LyC photons is significantly influenced by the gas covering fraction, which is the fraction of the ISM that is covered by optically thick hydrogen clouds capable of absorbing or scattering ionizing photons. Therefore, if the ISM is characterized by regions of lower hydrogen column density ($N_{HI}< 10^{17}~\rm cm^{-2}$), i.e. it has a covering fraction $< 1$,  LyC photons have a higher chance to escape the galaxy through the optically thin channels. 

The covering fraction of the neutral gas is usually probed by UV, not saturated absorption lines produced by ions in lower ionization states. We will focus on absorption lines emitted by ions with lower ionization potentials (LIS), which probe the dense neutral gas in the galaxy. These lines include Si II at 1260 \AA, 1304 \AA, and 1526 \AA, C II at 1334 \AA, O I at 1302 \AA, Fe II at 1608 \AA, and Mg II at 2796 \AA and 2803 \AA. Once the intensity of the absorption lines is determined, we will apply the Apparent Optical Depth method \citep{Savage&Sembach1991} to lines emitted by the same ion to verify their level of saturation. We will use non-saturated lines to determine the gas covering fraction by analyzing the residual intensity at the line center of the absorption feature \citep{Chisholm+2018, Steidel+2018, Gazagnes+2020, Saldana-Lopez+2022}. Our analysis will also include the H I Lyman series absorption lines, which probe the less dense gas \citep{Saldana-Lopez+2022, Jaskot+2024a, Jaskot+2024b, Flury+2024}. The strength of the unsaturated lines will allow us to determine the column density of neutral gas, providing insights into the ISM conditions that facilitate the escape of LyC photons. 

\vspace{1mm}
\noindent
\underline{Requirements}: For points 3) and 4) our requirement is to be able to detect the continuum. This is in fact fundamental to be able to detect absorption lines. The requirements described in the other points enable this science objective as well.
\\

\noindent 5) \textbf{SFR surface density}. In compact galaxies, the high SFR surface density is able to induce strong stellar winds and feedback. These phenomena remove neutral hydrogen, reducing the optical depth and allowing LyC photons to escape through carved optically thin channels. Additionally, the smaller physical size of compact galaxies implies that LyC photons have a shorter path to travel before escaping the galaxy \citep{Heckman+2011, Borthakur+2014, Jaskot+2013, Izotov+2016a, Izotov+2016b, Izotov+2018a, Izotov+2018b}.

We will determine the galaxy size using optical photometry, which allows us to measure the extent and structure of the galaxy in visible light. Furthermore, we will use the H$\alpha$ emission line to calculate the SFR \citep[e.g.,][]{Theios+2019}. By combining these two quantities—the galaxy size and the SFR derived from the Balmer lines or the UV continuum—we will compute the $\rm \Sigma SFR$. This parameter provides insight into the intensity of star formation per unit area, which is related to star-forming activity and the feedback processes that can influence the escape of LyC photons. The determination of the SFR requires the use of emission lines that are stronger than He II lines. Therefore, we are confident that the criteria described in points 1) and 2) are also suitable for determining the SFR.

\vspace{1mm}
\noindent
\underline{Requirements}: Our major progress is represented by reaching  $m_{UV}$ magnitudes fainter than 25 ABmag. We expect our galaxies to be metal poor (because they are low mass galaxies) and therefore their optical flux (at $\sim$ 5000 \AA) $\sim$ 10 times fainter than that at 1500 \AA. Therefore, we require the optical filter to have a depth of approximately ABmag $\sim$ 27.5.
\\

\noindent
We highlight that the calibrations of \fesc\ and its indirect indicators will be performed on galaxies at $z< 0.1$. This redshift range is required because the Ultraviolet Explorer Survey (UVEX), the survey we will use as our primary data source (see subsection \ref{subsec:sample}), can reach the faint magnitudes we require ($M_{UV}\sim-12.7$) within this redshift range.

\section{Physical parameters}
\label{sec:phys}

Our science objective is to derive the escape fraction \fesc\ of ionizing photons in low mass, faint galaxies at $z<0.1$, and calibrate indirect indicators on the derived \fesc. Therefore, the main physical parameter that we aim to obtain is the UV luminosity (absolute magnitude) of the observed galaxies. Specifically, our major goal is to reach absolute ab magnitudes in the UV as faint as \Muv $\sim -13$, which corresponds to  apparent \Muv $\sim 25$ ABmag (at $\sim 1500$ \AA). 

\subsection{Sample size}
\label{subsec:sample}
Calibrations of \fesc\ and its indirect indicators using brighter galaxies from the LzLCS+ sample \citep{Flury+2022a, Flury+2022b} have revealed significant scatter in the relationships between LyC emission and individual galaxy properties. This suggests that relying on a single indicator may be insufficient, and that a combination of properties could more effectively predict \fesc. Building on this idea, \citet{Jaskot+2024a} and \citet{Jaskot+2024b} conducted a multivariate analysis of the LzLCS+ galaxies, ranking the indicators based on their predictive power. Their results show that the most reliable predictors of \fesc\ are the equivalent width (EW) of Lyman-series absorption lines and UV dust attenuation, both of which trace line-of-sight absorption by neutral hydrogen and dust.

In this science case, we propose to perform a similar multivariate analysis on faint galaxies. 
To better map the distribution of the galaxy properties described in Section \ref{sec:objective}, we will define four bins for each parameter. For our major progress, we will require a minimum of 10 objects per bin. This implies a total of $\sim$ 40,000 objects, which represents a $\sim 20$-fold increase statistical power compared with the analysis performed on LzLCS+ galaxies. This improvement will enable a comprehensive exploration of the parameter space, not only through parameter projections, but also by examining how each individual parameter correlates with \fesc.

\subsection{Why do we need the Habitable World Observatory?}

Current and near-future facilities are not able to reach the sensitivity required to carry out this science case.
DESI and TMT can directly observe LyC at $z\sim3.5$ and $z\sim2.7$, respectively. However, at higher redshift, the IGM absorption becomes more and more significant, and requires the observations of independent sightlines to minimize the effect of correlated Ly$\alpha$ forest absorption \citep{Scarlata+2025}. By focusing on local galaxies, HWO will be able to avoid IGM-related uncertainties on the \fesc\ measurements. 

The Roman Space Telescope  will not be able to observe LyC directly, but it will be able to capture the UV indirect indicators at $z < 2.3$.  However, follow ups of LyC on the Roman targets is unfeasible. Roman will have a sensitivity of about 28.5 at $\sim3000$ \AA. Even in the best case scenario of a dust poor, young stellar population, this implies a faint UV ABmag of $\sim$ 25.5 at 1500 \AA. This, in turn, implies  an ABmag in LyC of 27.5 (assuming 100\% \fesc). This limit is beyond the reach of current facilities that are able to capture LyC at such redshifts.

We anticipate that suitable candidates for follow-up with the Habitable Worlds Observatory will be among the targets observed with the Ultraviolet Explorer (UVEX). UVEX will explore the full-sky and observe $\sim 10 – 200$ million galaxies with masses between $10^6$ and $10^9~ M_{\odot}$ at $z < 0.1$ \citep{Kulkarni+2021}. Note that galaxies with $M < 10^9~M_{\odot}$ in the local Universe have \Muv $> -19$, which is consistent with our threshold magnitude of \Muv = -18. Moreover, UVEX is expected to reach an absolute absolute FUV ABmag of $\sim$ -12.6 (\Muv $\sim$ 25.7) at this redshift (priv. comm.), which is in alignment with our science goal. With a planned launch around 2030, UVEX will provide ample time to identify and follow up promising candidates for HWO observations.

\begin{table*}[h]
\caption{Progress in galaxy sample and UV luminosity.}
\small
    \centering
    \begin{tabular}{c|c|c|c|c}
        \hline
        \textbf{Physical Parameter} & \textbf{State of the Art} & \textbf{Incremental Progress} & \textbf{Substantial Progress} & \textbf{Major Progress} \\
        & & \textbf{(Enhancing)} & \textbf{(Enabling)} & \textbf{(Breakthrough)} \\
        \hline
        \textbf{Number of galaxies in} & $\sim 90$ & $\sim 400$ & $\sim 2000$ & $\sim 40000$ \\
        \textbf{the studied sample} & & ($\sim 2$-fold S/N & ($\sim 5$-fold S/N & ($\sim 20$-fold S/N \\
        & & increase) & increase) & increase) \\
        \hline
        UV luminosity & $L \sim L^*$ & $L \sim 0.5L^*$ & $L \sim 1/10L^*$ & $L \sim 1/100L^*$ \\
        (\textit{in} $L^*$ \textit{units}) & ($M_{UV} \sim -18$) & ($M_{UV} \sim -17.2$) & ($M_{UV} \sim -16.5$) & ($M_{UV} \sim -13$) \\
        \hline
        Apparent $m_{UV}$  & 20 & 21 & 22.5& 25 \\
        at $z \sim0.1$ &  &  &  &  \\
        \hline
    \end{tabular}
    \label{tab:galaxy_progress}
\end{table*}

\section{Description of the observations}

\newcolumntype{L}[1]{>{\raggedright\arraybackslash}p{#1}}
\renewcommand{\arraystretch}{1,5} 

\begin{table*}[h]
\caption{Overview of observation requirements and progress levels.}
    \small 
    \centering
    \begin{tabularx}{\textwidth}{L{3.cm}|c|c|c|c}
        \hline
        \textbf{Observation Requirement} & \textbf{State of the Art} & \textbf{Incremental Progress} & \textbf{Substantial Progress} & \textbf{Major Progress} \\
        \hline
        LyC imaging depth & 27.5 ABmag & 28.5 ABmag & 30 ABmag & 32.5 ABmag \\
        \hline
        Spectral depth at $\sim$1600~\AA{} (wavelength of faintest He II) & 
        \begin{tabular}[c]{@{}c@{}}S/N continuum = 10 \\ (based on CLASSY galaxies, \\ where continuum level $\sim$ \\ $2 - 10 \times 10^{-15}$ erg/s/cm$^2$/\AA{})\end{tabular} & 
        S/N = 30 & S/N = 50 & S/N = 100 \\
        \hline
        Spectral resolution & \begin{tabular}[c]{@{}c@{}}R = 2,000 \\ (G140L for LzLCS)\end{tabular} &  &  & R = 10,000 \\
        \hline
    \end{tabularx}
    
    \label{tab:observation_requirements}
\end{table*}

\noindent 1) \textbf{Escape fraction of ionizing photons}. We need photometric far UV filters that are able to reach ABmag $\gtrsim 32.5$ at wavelengths $< 1000$ \AA. 
We do not need high resolution to measure \fesc\ in our targets, therefore we plan on determining \fesc\ using photometry.  Galaxies at the knee of the UV luminosity function at $z\sim0$ have \Muv $\sim-18$ (which corresponds to $m_{UV}\sim20$), which approximately corresponds to the faintest UV magnitude where \fesc\ has been probed in the local Universe. Our ultimate aim is to reach UV Luminosities  of $\sim$ 1/100 $L^*$. These are the faintest magnitudes where both \ro\ and \xion\ have been probed. For a typical non-ionizing-to-ionizing flux ratio of 10 and an assumed \fesc\ of 1 \%, a galaxy’s flux at the LyC wavelengths ($< 1000$ \AA) can be up to 1000 times fainter than the flux above the Lyman break. Therefore, our ultimate aim is to reach $\gtrsim 32.5$  at wavelength $< 1000$ \AA. An undetected LyC emission would be able to only put upper limits on the LyC escape fraction. However, the multivariate analysis we plan to perform incorporates both detections and non-detections. Therefore, LyC non-detections will still be informative. 
\\

\noindent 2) \textbf{Star formation rate surface density and dust extinction}: we need optical filters that reach ABmag $\sim27.5$.
Alongside a far-UV filter to capture LyC emission, we also need redder filters that extend into the optical wavelengths. We expect the spectra of our targets to be blue (since the galaxies will have low mass and therefore will be metal poor). Therefore, we anticipate the flux at optical wavelengths to be $\sim$ 10 times fainter than the UV flux. Therefore, we require the optical filter to have a depth of approximately ABmag $\sim27.5$. These are crucial for determining the compactness, and therefore the star formation rate surface density of our targets. From the LyC images of our low-mass galaxies at the lowest redshifts,  we will also be able to identify the spatial locations of LyC escape and correlate them with the properties of specific star-forming regions. This effort will be complementary to the HWO science case currently led by Dr. Xinfeng Xu, providing a comprehensive understanding of the mechanisms driving LyC escape in faint galaxies.
\\

\noindent 3) \textbf{Ionization state}: we require spectral coverage from 1000 to 5100 \AA\ restframe and spectral resolution $R \gtrsim 10000$. Our science goal requires us to cover wavelengths from $\sim1000$ \AA\ to $\sim5100$ \AA\ rest frame. This coverage will capture the indirect indicators listed in Step 2, i.e.  Si II at 1260 \AA\ and 1304 \AA, and 1526 \AA, C II at 1334 , O I at 1302 \AA, Fe II at 1608 \AA, and Mg II at 2796 \AA, and 2803 \AA, C III] at 1907,9  \AA, C IV at 1548 \AA, [O II] at 3726,9 \AA, and [O III] at 5007 \AA. 
For our major progress we aim at increasing the S/N (spectral depth) close to the expected faintest line by a factor 10 compared to what current low mass galaxies have (see CLASSY Survey).
We aim for a resolving power 10 times larger than what was used for the study of LzLCS galaxies, i.e., $R \sim 10000$. This will allow distinguishing features separated by only  30 $km~s^{-1}$. This is needed to study the ISM hydrogen distribution through the MgII double peak profile.
\\

\noindent  4) \textbf{Covering fraction}: we require to detect the continuum with a S/N $>$ 10 (which represents the state of the art) and spectral resolution $R \gtrsim 10000$. Detecting the S/N with a high S/N is a crucial requirement to be able to detect the UV absorption lines, even the weakest ones (such as C IV and Si IV). The high spectral resolution is needed to be able to separate the ISM absorption lines from the close stellar features (e.g., Mg II lines). 
\\

\noindent If our observations succeed, we will be able to pinpoint the role of faint galaxies in the reionization of the universe, potentially identifying them as the key drivers of this cosmic transformation. This will allow us to definitively determine whether faint or bright galaxies were the dominant contributors to reionization, resolving one of the most fundamental questions in early universe studies. 
\\

{\bf Acknowledgements.} Based on observations with the NASA/ESA Hubble Space Telescope obtained at the Space Telescope Science Institute, which is operated by the Association of Universities for Research in Astronomy, Incorporated, under NASA contract NAS5-26555. Support for
Program number 16734 was provided through a grant from the
STScI under NASA contract NAS5-26555. We acknowledge
the support from the LBT-Italian Coordination Facility for the
execution of observations, data distribution, and reduction.
The LBT is an international collaboration among institutions in
the United States, Italy, and Germany. The LBT Corporation
partners are The University of Arizona on behalf of the
Arizona university system; Istituto Nazionale di Astrofisica, Italy; LBT Beteiligungsgesellschaft, Germany, representing the Max-Planck Society, the Astrophysical Institute Potsdam, and Heidelberg University; and The Ohio State University, and
The Research Corporation, on behalf of The University of
Notre Dame, University of Minnesota and University of
Virginia.

\bibliography{author.bib}

\end{document}